\documentstyle[preprint,aps]{revtex}

\input psfig.sty
\draft

\begin{document}
\title{Neutrino-pair bremsstrahlung from nucleons in the condensed pion
field: revisited}
\author{L. B. Leinson}
\address{Institute of Terrestrial Magnetism, Ionosphere and Radio Wave\\
Propagation RAS, 142090 Troitsk, Moscow Region, Russia}
\maketitle

\begin{abstract}
At temperatures less than a few $MeV$, the efficiency of 
the neutrino-pair bremsstrahlung from nucleons interracting with the condensed pion field 
is much less than previously estimated by other authors. 
The physical reason for this is a periodic structure of the developed pion 
condensate. The repeated interactions of a nucleon with the periodic pion field 
modify the nucleon spectrum, which is split into two bands. The energy gap 
between the bands is about a few $MeV$ or larger, even if the amplitude of the 
pion field is small compared with the pion mass.
\end{abstract}


\widetext

\section{Introduction}

The theory of the neutrino-pair radiation from nucleons in the $\pi ^{0}$-
condensed nucleon matter of neutron stars was developed in \cite%
{Voskresensky} under the basic simplification that the amplitude of the pion
field is small with respect to the pion mass, $\left| {\bf \varphi }\right|
^{2}\ll m_{\pi }^{2}$, so that the nucleon-pion interaction may be
considered as a perturbation, and the matrix element of the reaction is
given by the following diagrams,

\psfig{file=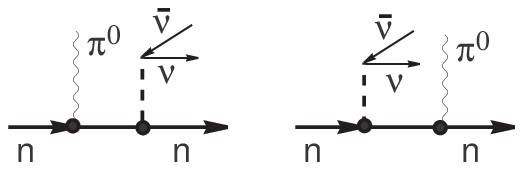} where the wavy line represents the external field of
the pion condensate, and the initial and final nucleon states are the plane
waves. The conclusion has been made that the corresponding neutrino energy
losses depend on temperature as $T^{6}$.

What we demonstrate in this letter is that the above result is valid only in
the more strong limit, $\left| {\bf \varphi }\right| ^{2}\ll T^{2}$, which
is never fulfilled in the case of developed condensate field. In the more
realistic case, $\left| {\bf \varphi }\right| ^{2}\gg T^{2}$, the process is
suppressed exponentially. The physical reason for this is a periodic
structure of the developed pion condensate. The repeated interactions of a
nucleon with the periodic pion field modify the nucleon spectrum, which is
split into two bands, and the energy gap between the bands is about a few $%
MeV$ or larger, even if the amplitude of the pion field is small compared
with the pion mass. Consequently, at temperatures less than a few $MeV$, at
which the above reactions has been thought to be important, the efficiency
of the process is much less than previously estimated. The analogous
mechanism of suppression of the neutrino-pair radiation was considered for
the electron bremsstrahlung in the crystalline crust of the neutron star %
\cite{Pethik}, and for neutrino emission from the bubble phase of the
stellar nuclear matter \cite{L1993}.

In the following we use the system of units $\hbar =c=1$.

\section{Nucleon states in the pion field}

By assuming a pseudovector nucleon interaction with the pion field ${\bf %
\varphi =}\left( \varphi _{1},\varphi _{2},\varphi _{3}\right) $ the
effective nucleon Hamiltonian can be written as 
\begin{equation}
{\cal H}_{nr}=-\frac{1}{2M^{\ast }}{\bf \nabla }^{2}{\bf +}\frac{f}{m_{\pi }}%
\left( {\bf \sigma }\nabla \right) \left( {\bf \varphi \tau }\right)
\label{Hnr}
\end{equation}%
where the spin, $\sigma _{i}$, and isospin, $\tau _{i}$, operators are given
by the Pauli matrices; $M^{\ast }$ is the effective nucleon mass; and $%
f=0.988$ is the pion-nucleon coupling constant.

To examine the nucleon spectrum in the presence of the $\pi ^{0}$
condensate, let us consider a simple model with the condensed classical pion
field ${\bf \varphi =}\left( 0,0,\varphi _{0}\right) $ of the form of a
standing wave \cite{Migdal} 
\begin{equation}
\varphi _{0}\left( {\bf r}\right) =\sqrt{\frac{2}{3}}a\left( \sin kx+\sin
ky+\sin kz\right)   \label{fi}
\end{equation}%
with a real amplitude $a$, so that $a^{2}=\left\langle {\bf \varphi }%
^{2}\right\rangle .$ In this case we have%
\begin{equation}
{\cal H}_{nr}=-\frac{1}{2M^{\ast }}{\bf \nabla }^{2}{\bf +}\sqrt{\frac{2}{3}}%
f\frac{a}{m_{\pi }}k\left( {\bf \sigma R}\right) \tau _{3}
\end{equation}%
with%
\begin{equation}
{\bf R}\left( {\bf r}\right) =\left( \cos kx,\cos ky,\cos kz\right) .
\label{R}
\end{equation}%
Thus the nucleon quasi-particle moves in the three-dimensional periodic
potential%
\begin{equation}
U\left( {\bf r}\right) =\sqrt{\frac{2}{3}}f\frac{a}{m_{\pi }}k\left( {\bf %
\sigma R}\right) \tau _{3},
\end{equation}%
which can be recast as the sum over reciprocal lattice vectors%
\begin{equation}
U\left( {\bf r}\right) {\bf =}\sqrt{\frac{2}{3}}f\frac{a}{m_{\pi }}%
kt_{3}\sum_{{\bf K}}\left( {\bf \sigma R}_{{\bf K}}\right) \exp \left( i{\bf %
Kr}\right) .  \label{Ur}
\end{equation}%
Here $t_{3}=\pm 1$ for protons and neutrons respectively. The matrix element 
\begin{equation}
{\bf R}_{{\bf K}}=\frac{1}{V}\int_{V}d^{3}r{\bf R}\left( {\bf r}\right) \exp
\left( i{\bf Kr}\right) 
\end{equation}%
is given by the integral over the volume of the elementary cell%
\begin{equation}
V=\left( \frac{2\pi }{k}\right) ^{3}.
\end{equation}%
One can easily find that there are only six non-zero reciprocal lattice
vectors for which the matrix element does not vanish, ${\bf R}_{{\bf K}}\neq
0$. The direct calculation yields%
\begin{equation}
{\bf R}_{{\bf K}}=\frac{1}{2}{\bf n}_{{\bf K}},
\end{equation}%
where the unit vector ${\bf n}_{{\bf K}}$ has the following components
\begin{eqnarray}
{\bf n}_{{\bf K}} &=&{\bf n}_{{\bf -K}}=\left( 1,0,0\right) \ \ \ \ \text{if
\ \ \ }{\bf K}=\left( \pm k,0,0\right)   \nonumber \\
{\bf n}_{{\bf K}} &=&{\bf n}_{{\bf -K}}=\left( 1,0,0\right) \ \ \ \ \text{if
\ \ \ }{\bf K}=\left( \pm k,0,0\right)   \label{UK} \\
{\bf n}_{{\bf K}} &=&{\bf n}_{{\bf -K}}=\left( 0,0,1\right) \ \ \ \ \text{if
\ \ \ }{\bf K}=\left( 0,0,\pm k\right)   \nonumber
\end{eqnarray}%
Thus we can write%
\begin{equation}
U\left( {\bf r}\right) {\bf =}U_{k}t_{3}\sum_{{\bf K}}\left( {\bf \sigma n}_{%
{\bf K}}\right) \exp \left( i{\bf Kr}\right) ,
\end{equation}%
where 
\begin{equation}
U_{k}=\frac{1}{2}\sqrt{\frac{2}{3}}fa\frac{k}{m_{\pi }}  \label{Uk}
\end{equation}

By assuming that the periodic potential $U\left( {\bf r}\right) $ is small
with respect to the nucleon energy, near the Fermi surface, the nucleons can
be considered as almost free quasi-particles of the effective mass $M^{\ast
} $. Then the crystal potential has the most effect on the nucleon states
for which the free particle energies $\varepsilon _{{\bf p}}$ and $%
\varepsilon _{{\bf p-K}}$ are almost equal for some reciprocal lattice
vector ${\bf K}$, or, equivalentely, if the component of the quasi-particle
momentum in the direction of the reciprocal lattice vector is close to $%
p_{\parallel }\simeq K/2$. The latter condition means that, in the above
approximation, the summation over reciprocal lattice vectors can be
restricted by the values 
$\left| {\bf K}\right| <2p_{F}\text{.}$
Typically the pion condensation appears at $k\simeq \left( 1.4\div
1.7\right) m_{\pi }$, and the condition (\ref{Kmax}) means the summation
over all of the six reciprocal lattice vectors as given by Eq. (\ref{UK}).

At $p_{\parallel }\simeq K/2$ one has $\varepsilon _{{\bf p}}\simeq
\varepsilon _{{\bf p-K}}$. In this case, up to the leading terms, the
equation of motion of the nucleon has the form%
\begin{equation}
\left[ E+\frac{1}{2M}{\bf \nabla }^{2}{\bf +}t_{3}U_{k}e^{i{\bf Kr}}\left( 
{\bf \sigma n}_{{\bf K}}\right) \right] \psi =0,
\end{equation}%
and the nucleon wave function can be written as a superposition of two plane
waves%
\begin{equation}
\psi \left( {\bf r,\sigma }\right) =\left[ A\left( {\bf p,}\sigma \right)
\exp \left( i{\bf pr}\right) +B\left( {\bf p,}\sigma \right) \exp \left( i%
{\bf pr-}i{\bf Kr}\right) \right] ,
\end{equation}%
where the functions $A\left( {\bf p,}\sigma \right) $ and $B\left( {\bf p,}%
\sigma \right) $ obey the following set of equations%
\begin{equation}
\left( E-\varepsilon _{{\bf p}}\right) A\left( {\bf p,}\sigma \right) {\bf +}%
t_{3}\left( {\bf \hat{\sigma}n}_{{\bf K}}\right) U_{k}B\left( {\bf p,}\sigma
\right) =0,  \label{A}
\end{equation}%
\begin{equation}
\left( E-\varepsilon _{{\bf p-K}}\right) B\left( {\bf p,}\sigma \right)
+t_{3}\left( {\bf \hat{\sigma}n}_{{\bf K}}\right) U_{k}A\left( {\bf p,}%
\sigma \right) =0,  \label{B}
\end{equation}%
$\allowbreak $which has a non-trivial solution if 
\begin{equation}
\left( E-\varepsilon _{{\bf p-K}}\right) \left( E-\varepsilon _{{\bf p}%
}\right) {\bf -}U_{k}^{2}=0.
\end{equation}%
As it follows from this dispersion equation, the energy eigenvalues, are
devided into two bands, 
\begin{equation}
E_{{\bf p}}^{\pm }=\frac{\varepsilon _{{\bf p}}+\varepsilon _{{\bf p-K}}}{2}%
\pm \sqrt{\left( \frac{\varepsilon _{{\bf p}}-\varepsilon _{{\bf p-K}}}{2}%
\right) ^{2}+U_{k}^{2}},  \label{Epm}
\end{equation}%
shown in Fig. 1, where the upper and the lower band are denoted as ${\bf +}$
and $-$ respectively.

\psfig{file=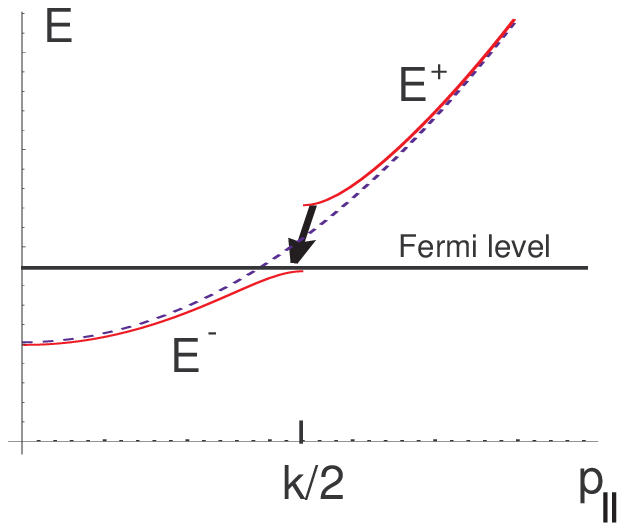}{Fig. 1 Red solid line represents the nucleon spectrum as given 
by Eq, (10). Dashed line corresponds to a free nucleon.  
The arrow shows a possible nucleon transition accompanied by the neutrino-pair 
emission.} $\allowbreak $

Near $\varepsilon _{{\bf p}}\simeq \varepsilon _{{\bf p-K}}$, the nucleon
spin function can be taken as the eigenstate of ${\bf \hat{\sigma}n}_{{\bf K}%
}${\bf :} 
\begin{equation}
\left( {\bf \hat{\sigma}n}_{{\bf K}}\right) \chi _{\lambda }=\lambda \chi
_{\lambda },\ \ \ \lambda =\pm 1,
\end{equation}%
Thus, from Eqs. (\ref{A}), (\ref{B}) we obtain the following normalized wave
function for the upper band%
\begin{equation}
\psi _{{\bf p,\lambda }}^{+}\left( {\bf r,\sigma }\right) =\left[ u_{{\bf p}%
}\exp \left( i{\bf pr}\right) {\bf -}t_{3}v_{{\bf p}}\left( {\bf \hat{\sigma}%
n}_{{\bf K}}\right) \exp \left( i{\bf pr-}i{\bf Kr}\right) \right] \chi
_{\lambda }\left( \sigma \right) ,  \label{p}
\end{equation}%
while for the lower band we have%
\begin{equation}
\psi _{{\bf p,\lambda }}^{-}\left( {\bf r,\sigma }\right) =\left[ v_{{\bf p}%
}\exp \left( i{\bf pr}\right) {\bf +}t_{3}u_{{\bf p}}\left( {\bf \hat{\sigma}%
n}_{{\bf K}}\right) \exp \left( i{\bf pr-}i{\bf Kr}\right) \right] \chi
_{\lambda }\left( \sigma \right) .  \label{m}
\end{equation}%
The ''coherence factors'' are given by%
\begin{equation}
u_{{\bf p}}^{2}=\frac{1}{2}\left( 1+\frac{\xi _{{\bf p}}}{\zeta _{{\bf p}}}%
\right) ,\ \ \ \ v_{{\bf p}}^{2}=\frac{1}{2}\left( 1-\frac{\xi _{{\bf p}}}{%
\zeta _{{\bf p}}}\right) ,\ \ \ u_{{\bf p}}v_{{\bf p}}=\frac{\left|
U_{k}\right| }{2\zeta _{{\bf p}}},
\end{equation}%
with 
\begin{equation}
\xi _{{\bf p}}=\frac{\varepsilon _{{\bf p}}-\varepsilon _{{\bf p-K}}}{2},\ \
\ \ \ \ \zeta _{{\bf p}}=\sqrt{\xi _{{\bf p}}^{2}+U_{k}^{2}}.
\end{equation}%
In contrast to the ordinary plane waves, in the nucleon wave functions (\ref%
{p}), (\ref{m}), the nucleon interaction with the condensed pion field is
''exactly'' taken into account\footnote{%
In our context we use term ''exactly'' to stress that the energy spectrum (%
\ref{Epm}) can be obtained only by summation of the perturbation series to
all orders.}. As it was expected the spectrum of the corresponding nucleon
states is of the band-like structure similar to that for electrons in a
cristalline metal. As given by Eq. (\ref{Epm}), the energy gap in the
nucleon spectrum is $\Delta =2U_{k}$. As will be shown in the next section,
even in the case of a small amplitude of the pion field, $T\ll a\ll m_{\pi }$%
, the energy gap, $U_{k}\gg T$, exponentially suppresses the neutrino-pair
bremsstrahlung from the nucleons, and the neutrino energy losses in this
process are much less than estimated before in \cite{Voskresensky}.

\section{Neutrino energy losses}

When, in the nucleon wave functions, the condensed pion field is ''exactly''
taken into account the neutrino-pair bremsstrahlung%
\begin{equation}
n\rightarrow n+\nu +\bar{\nu}  \label{br}
\end{equation}
can be described by the following diagram,

\psfig{file=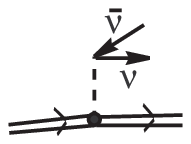} where the lepton pair is radiated by the nucleon
undergoing a transition between its eigen-states in the pion field. If the
initial and final nucleon states are in the same band, the radiation is
kinematically forbidden. Indeed, in this case the four-momentum $q=\left(
q_{0},{\bf q}\right) $ transferred from the nucleon is space-like\footnote{%
Since the group velocity of the nucleon is $\nabla _{{\bf p}}E<1$, the
energy difference between nucleon states $q_{0}=\left| {\bf q}\right| \nabla
_{{\bf p}}E$ is less than $\left| {\bf q}\right| $.}, $q_{0}<\left| {\bf q}%
\right| $, while the total momentum of the lepton pair is a time-like, $%
q_{0}>\left| {\bf q}\right| $. Consequently it is impossible simultaneously
to conserve the energy and momentum in the reaction. However, even for small
momentum transfers, there exists a finite energy difference between
different bands, so the neutrino radiation is kinematically allowed for the
nucleon transitions shown by the arrow in Fig. 1.

We now examine the process in which the neutron quasi-particle $\left(
t_{3}=-1\right) $ makes a transition from the upper band to the lower one.
The neutrino interaction with the non-relativistic neutrons is of the form%
\footnote{%
Since our goal is to prove the exponential suppression of the process, we
omit the correction to the weak vertex caused by NN correlations, which is
known to suppress the neutrino energy losses up to 100 times \cite%
{Voskresensky}.}%
\begin{equation}
\hat{H}=-\frac{G_{F}}{2\sqrt{2}}\,\left( \,J_{0}l_{0}-g_{A}\,{\bf J}{\bf l}%
\right) ,
\end{equation}%
where $G_{F}$ is the Fermi constant, and $g_{A}=1.26$. The vector current of
the non-relativistic neutron has only the time component, $J_{0}\equiv \hat{%
\Psi}^{\dagger }\hat{\Psi}$, while the axial-vector current is given by its
space components, ${\bf J}\equiv \hat{\Psi}^{\dagger }{\bf \sigma }\hat{\Psi}
$. The neutral weak current of neutrinos is of the standard form 
\begin{equation}
l_{\mu }=\bar{\Psi}_{\nu }\gamma _{\mu }\left( 1-\gamma _{5}\right) \Psi
_{\nu },
\end{equation}%
We consider the total energy which is emitted into neutrino pairs per unit
volume and time. By Fermi's ''golden'' rule we have%
\begin{eqnarray*}
Q &=&\sum_{\nu }\int \frac{d^{3}pd^{3}p^{\prime }}{(2\pi )^{6}}\int \frac{%
d^{3}q_{1}}{2q_{1}^{0}(2\pi )^{3}}\frac{d^{3}q_{2}}{2q_{2}^{0}(2\pi )^{3}}%
2\pi \delta \left( E_{{\bf p}}^{+}-E_{{\bf p}^{\prime }}^{-}-q^{0}\right) \\
&&\;\times q^{0}\sum_{{\bf K}}\sum_{\lambda }\overline{\left| {\cal M}%
_{fi}\right| ^{2}}f\left( E_{{\bf p}}^{+}\right) \left( 1-f\left( E_{{\bf p}%
^{\prime }}^{-}\right) \right) ,
\end{eqnarray*}%
The integration goes over the phase volume of neutrinos and antineutrinos of
total energy $q^{0}=q_{1}^{0}+q_{2}^{0}$ and total momentum ${\bf q=q}_{1}+%
{\bf q}_{2}$. The symbol $\sum_{\nu }$\ \ indicates that summation over the
three neutrino types has to be performed. The square of the matrix element
of the reaction (\ref{br}) summed over spins of initial and final particles
has the following form%
\begin{equation}
\sum_{\lambda }\overline{\left| {\cal M}_{fi}\right| ^{2}}%
=G_{F}^{2}g_{A}^{2}u_{{\bf p}}^{2}v_{{\bf p}}^{2}\left( 2\pi \right)
^{3}\delta \left( {\bf p}-{\bf p}^{\prime }\right) \delta _{\mu i}\delta
_{\nu j}\left( \delta _{ij}-\delta _{i3}\delta _{j3}\right) 
\mathop{\rm Tr}%
l_{\mu }l_{\nu }
\end{equation}%
Here we take into account that, in the degenerate case $T\ll \mu $, one can
neglect a small momentum of the neutrino par, $\left| {\bf q}\right| \sim T%
{\bf \ll }\left| {\bf p}_{F}\right| $, by making the following replacement $%
\delta \left( {\bf p}-{\bf p}^{\prime }-{\bf q}\right) \simeq \delta \left( 
{\bf p}-{\bf p}^{\prime }\right) $. The distribution function of initial
nucleon as well as blocking of its final states are taken into account by
the Pauli blocking-factor $f\left( E^{+}\right) \left( 1-f\left(
E^{-}\right) \right) $, where $f$ is the Fermi distribution function%
\begin{equation}
f\left( E\right) =\left( \exp \left( \frac{E-\mu }{T}\right) +1\right) ^{-1}.
\end{equation}%
We assume that neutrinos can escape freely from the matter.

By inserting $\int d^{4}q\delta ^{\left( 4\right) }\left(
q-q_{1}-q_{2}\right) =1$ in this equation, and making use of the Lenard's
integral \ 
\begin{eqnarray}
&&\int \frac{d^{3}q_{1}}{2q_{1}^{0}}\frac{d^{3}q_{2}}{2q_{2}^{0}}\;\delta
^{\left( 4\right) }\left( q-q_{1}-q_{2}\right) {\rm Tr}\left( l_{\mu }l_{\nu
}^{\ast }\right)  \nonumber \\
&=&\frac{4\pi }{3}\left( q_{\mu }q_{\nu }-q^{2}g_{\mu \nu }\right) \Theta
\left( q_{\mu }^{2}\right) \Theta \left( q^{0}\right) ,  \label{Lenard}
\end{eqnarray}%
where $\Theta (x)$ is the Heaviside step function, we can write%
\begin{eqnarray*}
Q &=&\sum_{\nu }\sum_{{\bf K}}\frac{4\pi }{3}G_{F}^{2}g_{A}^{2}\int d^{4}q\
\Theta \left( q_{\mu }^{2}\right) \Theta \left( q^{0}\right) q^{0}\int \frac{%
d^{3}p}{(2\pi )^{9}}2\pi \delta \left( E_{{\bf p}}^{+}-E_{{\bf p}%
}^{-}-q^{0}\right) \\
&&\;f\left( E_{{\bf p}}^{+}\right) \left( 1-f\left( E_{{\bf p}}^{-}\right)
\right) \left( q_{\mu }q_{\nu }-q_{\lambda }^{2}g_{\mu \nu }\right) u_{{\bf p%
}}^{2}v_{{\bf p}}^{2}\delta _{\mu i}\delta _{\nu j}\left( \delta
_{ij}-\delta _{i3}\delta _{j3}\right)
\end{eqnarray*}%
Then the direct calculation of the integrals gives 
\begin{equation}
Q=\sum_{\nu }\sum_{{\bf K}}\frac{1}{60\pi ^{5}}G_{F}^{2}g_{A}^{2}M^{2}\frac{%
U_{k}^{2}}{K}\int_{2U_{K}}^{\infty }\frac{q_{0}^{6}}{\sqrt{%
q_{0}^{2}-4U_{k}^{2}}}\frac{dq_{0}}{e^{\frac{q_{0}}{T}}-1}.
\end{equation}%
Here the lower limit of integration corresponds to the minimal energy of the
neutrino pair radiated due to the nucleon transition from the upper to the
lower band.

According to Eq. (\ref{UK}), there are only six non-zero reciprocal lattice
vectors for which the matrix element ${\bf R}_{{\bf K}}$ does not vanish. By
performing summation over reciprocal lattice vectors and taking into account
three neutrino flavours, $\sum_{\nu }=3$, we obtain 
\begin{equation}
Q=\frac{3}{10\pi ^{5}}G_{F}^{2}g_{A}^{2}M^{2}U_{k}^{2}\frac{1}{k}%
\int_{2U_{k}}^{\infty }\frac{q_{0}^{5}}{\sqrt{q_{0}^{2}-4U_{k}^{2}}}\frac{%
dq_{0}}{e^{\frac{q_{0}}{T}}-1}.  \label{Q}
\end{equation}
The energy gap 
\begin{equation}
\Delta =2U_{k}=\sqrt{\frac{2}{3}}f\frac{a}{m_{\pi }}k
\end{equation}%
between the nucleon energy bands rapidly increases along with increasing of
the amplitude of the condensed pion field. For a developed condensate one
has $2U_{k}\gg T.$ In this low-temperature limit, typical for neutron stars,
the neutrino energy losses are exponentially suppressed. Indeed, in the case 
$2U_{k}/T\gg 1$, the main contribution to the integral, in Eq. (\ref{Q}),
comes from the vicinity of the lower limit where $x\simeq 2U_{k}/T$.
Therefore, in the case of developed pion condensate, we obtain the following
estimate 
\begin{equation}
Q_{{\rm low-temp}}\simeq \frac{24\sqrt{2}}{5\pi ^{5}}G_{F}^{2}g_{A}^{2}M^{2}%
\frac{1}{k}TU_{k}^{7}e^{-4U_{k}/T}.  \label{Qlt}
\end{equation}

In the limiting case of small amplitude of the pion field $2U_{K}\ll T$,
from Eq. (\ref{Q}) we have%
\begin{equation}
Q\simeq \frac{2\pi }{315}G_{F}^{2}g_{A}^{2}M^{2}f^{2}\frac{a^{2}}{m_{\pi
}^{2}}kT^{6},  \label{Qpert}
\end{equation}%
which reproduces the result obtained in \cite{Voskresensky} by the
perturbation theory. Notice that the energy losses (\ref{Qpert}) are
actually 3 times larger than that given in \cite{Voskresensky} due to
summation over 3 neutrino flavours.
\section{Conclusion}
To summarize, neutrino-pair bremsstrahlung from nucleons in the periodic field of the $\pi^0$ condensate is much less important than suggested by earlier estimates. The reason for this is that, at the temperatures typical for cooling neutron stars, the process is suppressed by band structure effects.

\end{document}